\documentclass[prd,longbibliography,superscriptaddress,showpacs,showkeys,floatfix,eqsecnum]{revtex4}
\usepackage{textcomp}
\usepackage{eurosym}
\usepackage{amsfonts}
\usepackage{array}
\usepackage{amsthm}
\usepackage{bm}
\usepackage{palatino}
\usepackage[colorinlistoftodos]{todonotes}
\usepackage{mathpazo}
\usepackage{supertabular}
\usepackage{changes}
\usepackage{subfig}
\usepackage{amssymb}
\usepackage{eurosym}
\usepackage{amsmath}
\usepackage{epsfig}
\usepackage{graphics}
\usepackage{color}
\usepackage{graphicx}
\usepackage[font={footnotesize,it}]{caption}
\usepackage[colorlinks=true,
            linkcolor=red,
            urlcolor=blue,
            citecolor=green]{hyperref}

\begin{document}
\newcommand{\ds}{\displaystyle}
\newcommand{\be}{\begin{equation}}
\newcommand{\en}{\end{equation}}
\newcommand{\bea}{\begin{eqnarray}}
\newcommand{\ena}{\end{eqnarray}}

\title{Evolving topologically deformed wormholes supported in the dark matter halo}

\author{Ali \"{O}vg\"{u}n}
\email{ali.ovgun@emu.edu.tr}
\homepage{http://www.aovgun.com}

\affiliation{Physics Department, Eastern Mediterranean University, Famagusta, North Cyprus 99628 via Mersin 10, Turkey}

\date{\today}
\begin{abstract}

In this paper, we construct an evolving wormhole in the dark matter halo. This work is relevant since matter has two components: (i) cosmological part (only time-dependent) and (ii) wormhole part (only space-dependent). In order to implement this, we use the Chaplygin gas as an equation of state for the cosmic part and Navarro-Frenk-White dark matter density profile as well as Thomas-Fermi (TF) profile in order to form a dark wormhole. The flare-out condition of wormhole is also satisfied by violating the null energy condition (NEC) for some specific values of quantities. Furthermore, we reveal more interesting results regarding how an topologically deformation parameter $\alpha$ affects the evolving wormhole sourced with some dark matter models based on the physically motivated shape function. 

\end{abstract}


\maketitle
%\tableofcontents

\section{Introduction}
A wormhole is a theoretical passage through spacetime, which connects different points and creates a short-cut for traveling in a spacetime. Einstein and Rosen initially elaborated on wormholes in 1935 using the theory of general relativity. Subsequently, the main contributions on traversable wormholes were done by Morris, Thorne, and Yurtsever in the 1980s \cite{Morris:1988cz,Morris:1988tu}. However, since then, no one has discovered a wormhole until now, it is completely based on theoretical research. 
Understanding the mysterious of the nature of the universe is a puzzle for humanity.

The existence of wormhole geometries is required to utilize the exotic matter, which is one the most debatable issue. In order to minimize the presence of exotic matter, one is required to use modified theories or extra sources. An alternative approach can be the utilization of thin-shell formalism, which was initially proposed by Visser. Moreover, another important problem is to find a stable wormhole against perturbations. A singularity-free system identifies a stable state and it prevents the wormhole from collapsing. Recently, the studies of wormhole solutions have gained great interest \cite{Kim,Roman:1992xj,Frolov:1993jq,Gorini:2004by,Visser:1997yn,Kim:1997jf,Mirza:2006vr,Arellano:2006ex,Lobo:2014zla,Lobo:2014fma,Arellano:2008xw,Kashargin:2008pk,Maeda:2009tk,Dzhunushaliev:2009yw,ElNabulsi:2009zzj,Bochicchio:2010df,Guendelman:2009pf,Guendelman:1991pc,Sajadi:2012zz,Bochicchio:2013hd,Zangeneh:2014noa,Tomikawa:2014wxa,Najafi:2014xua,Tomikawa:2015swa,Bahamonde:2016ixz,Kim:2018dnl,Cataldo:2008pm,Cataldo:2008ku,Cataldo:2013vca,Cataldo:2017ard,Sharif1,Sharif3,Sharif5,Bronnikov:2002rn,Bronnikov:2012ch,Bronnikov:2009na,Bhattacharya:2015oma,Mishra:2017yrh,Halilsoy:2013iza,Sakalli:2015taa,Ovgun:2015una,Sakalli:2015mka,Ovgun:2015sqa,Ovgun:2016ijz,Ovgun:2016ujt,Jusufi:2016eav,Ovgun:2017zao,Ovgun:2017jip,Jusufi:2017vta,Jusufi:2017mav,Richarte:2017iit,Jusufi:2018kmk,Myrzakulov:2015kda,Bhar:2014ooa,Rahaman:2014pba,Ovgun:2020yuv,Ovgun:2018oxk,Ovgun:2018prw,Ovgun:2018fnk,Ovgun:2018xys,Rahaman:2016jds, Xu:2020wfm,KordZangeneh:2020jio,Lobo:2020kxn,Ayuso:2020vuu,KordZangeneh:2020ixt,Garattini:2017jlq,Kuhfittig:2021fes}.

On the other hand, theoretical and observational cosmology struggles to resolve the source of inflation in the primitive universe and also the present cosmic accelerated expansion. Recent experiments show that some enigmatic force may cause the accelerated expansion, which is known as the dark energy. Besides, it also highlights the ambiguous power of cosmology \cite{Ovgun:2017iwg,Ovgun:2016oit}.

Naturally, we assume that a wormhole can lead to an apparent failure of locality on the background spacetime in the primitive universe. If there was a wormhole in the early universe, it would be inflated into at least the size of a human to allow time travel, with reference to this, the early Roman studied the enlargement of the wormhole by inflation \cite{Roman:1992xj}. It was proposed that amidst the inflation era, the wormhole had inflated. Furthermore, Kim extended Roman's idea and studied the cosmological properties in the Friedmann-Robertson-Walker cosmologies with a traversable wormhole \cite{Kim}. He has divided the content of matter into two parts:  (i) the cosmological part (only time-dependent) and (ii) the wormhole part (only space-dependent). Subsequently, he studied the behavior of the scale factor and the wormhole shape function in this context. After these seminal papers, the study of exact solutions under assorted scenarios was extensively studied in order to understand the sophisticated picture of cosmic evolution and wormhole. After the creation of this predicament, Cataldo et al., in a subsequent paper \cite{Cataldo:2008pm,Cataldo:2008ku,Cataldo:2013vca,Cataldo:2017ard}, studied various models of an FRW like cosmologies using the wormholes in different dimensions. He argued that it is possible to find normal matter wormhole in the universe, and the evolving wormhole metric can cause the acceleration of the universe.

In this paper, our main aim is to study evolving topologically deformed  wormhole supported in dark matter halo, especially the Navarro-Frenk-White (NFW) profile \cite{NFW}, which is a spatial mass distribution of dark matter as well as Thomas-Fermi(TF) profile, on the other hand, the cosmic part sourced with the Chaplygin gas (CG) \cite{Gorini:2004by}. Differently from the previous studies, we investigate the effect of the topologically deformation parameter $\alpha$ on the evolving wormhole.

The remainder of this paper is organized as follows: In section II, we study the evolution of the universe using the CG gas in the topologically deformed FRW spacetime. Then in section III, we construct a dynamic traversable wormhole by solving Einstein field equations in the background of the topologically deformed FRW spacetime. In section IV we will discuss the results.

\section{Topologically deformed Wormhole embedded in FRW cosmology}

In this section, we first consider the spacetime metric representing a dynamic traversable wormhole in a FRW universe as follows:

\begin{eqnarray} d s^{2} &=-e^{2 \Phi(r)}d t^{2}+R(t)^{2}\left(\frac{d r^{2}}{1-k r^{2}-\frac{b(r)}{r}}+r^{2} \alpha^2 d \Omega^{2}\right), \label{4dysme}\end{eqnarray}
where  $b$ is function of $r$ and known as a shape function, $\alpha$ is the solid angle deficit ($0< \alpha< 1$) as well as $\phi(r)$ is the lapse function. $R(t)$
is the scale factor of the universe. Moreover, $k$ is the curvature of spacetime with values $+1,0,-1$. Note that $R(t) \to constant$ the static Morris-Thorne wormhole is recovered.

To avoid the form of event horizon, these conditions should be satisfied: $(1-k r^{2}-\frac{b(r)}{r})>0$, and
$b(r_0)=r_0$ at the throat. % and the flaring out condition $(b-b'r)/b^2 \geq 0$, with $b'(r_0)<1$ at the throat. Note that a prime stands for differentiation with respect to $r$. 
Solving the Einstein field
equation, $G_{\mu \nu}=8\pi T_{\mu \nu}$ for the above metric and $\phi(r)=0$, we obtain the nonzero components
of the Einstein tensor with energy density $\rho$, radial pressure $P^r$ and laterial pressure $P^t$ reduce to 

\begin{eqnarray}
\rho=\frac{1}{8 \pi}\left[\frac{3\left(\dot{R}^{2}+k\right)}{R^{2}}+\frac{1}{R^{2}} \frac{b^{\prime}}{r^{2}}+\,{\frac {-{\alpha}^{2}+1}{\,{r}^{2}{\alpha}^{2} R^{2}}}
\right],
      \label{4rGott}  \\
P^r=\frac{1}{8 \pi}\left[-2 \frac{\ddot{R}}{R}-\frac{\left(\dot{R}^{2}+k\right)}{R^{2}}-\frac{1}{R^{2}} \frac{b}{r^{3}}-{\frac { \left( \alpha-1 \right)  \left( \alpha+1 \right)  \left( -k{
r}^{3}+ \left( {R}^{2}+1 \right) r-b \right) }{{r}^{2}{R}^{2}{\alpha}^
{2} \left( k{r}^{3}+b-r \right) }}
\right],  \label{4rGorr}   \\
P^t=\frac{1}{8 \pi}\left[-2 \frac{\ddot{R}}{R}-\frac{\left(\dot{R}^{2}+k\right)}{R^{2}}+\frac{1}{2 R^{2}}\left(\frac{b}{r^{3}}-\frac{b^{\prime}}{r^{2}}\right)\right].
             \label{4rGohhpp}
\end{eqnarray}
%where the perfect fluid with the energy-momentum tensor given by $T_{ij} = diag[\rho(t, r),-\tau (t, r), P(t, r), P(t, r)]$, where using
Note that radial tension $\tau=-P^r$ is equal to the negative radial pressure and $H=\dot{R}/R$ is the cosmological Hubble parameter. Moreover an overdot stands for differentiation with respect to $t$. Then we use the following ansatz for matter parts to separate field equations in two parts \cite{Kim}

\begin{eqnarray}\label{uno}
R^{2}(t) \rho(r, t) &=R^{2}(t) \rho_{c}(t)+\rho_{w}(r), \\ \label{uno1}
R^{2}(t) P^{r}(r, t) &=R^{2}(t) P_{c}(t)+P_{w}^{r}(r),\\ \label{uno2}
R^{2}(t) P^{t}(r, t) &=R^{2}(t) P_{c}(t)+P_{w}^{t}(r). 
\end{eqnarray}
Note that above equations depend on $R^2$, which shows that the wormhole affects the curvature.
 Furthermore, we use the subscripts $c$ and $w$ to refer the cosmic and wormhole parts. %The radial pressure should be $\tau_c=-P_c$, to make them isotropic.  
 With the help of ansatz Eq.s~\ref{uno}-\ref{uno2}, we rewrite  the Einstein
equations in two parts as follows

\begin{eqnarray} \label{S00}
R^{2}\left[8 \pi \rho_{c}-3\left(\frac{\dot{R}}{R}\right)^{2}-\frac{3 k}{R^{2}}\right]=\frac{b^{\prime}}{r^{2}}-8 \pi \rho_{w}+{\frac {-{\alpha}^{2}+1}{{r}^{2}{\alpha}^{2}}}=l,  \\ \label{S11}
R^{2}\left[8 \pi P_{c}+2 \frac{\ddot{R}}{R}+\left(\frac{\dot{R}}{R}\right)^{2}+\frac{k}{R^{2}}\right]=-\frac{b}{r^{3}}-8 \pi P_{w}^{r}-{\frac { \left( \alpha-1 \right)  \left( \alpha+1 \right)  \left( -k{
r}^{3}+r{R}^{2}-b+r \right) }{{r}^{2}{\alpha}^{2} \left( k{r}^{3}+b-r
 \right) }}
=m,  \\ \label{S22}
R^{2}\left[8 \pi P_{c}+2 \frac{\ddot{R}}{R}+\left(\frac{\dot{R}}{R}\right)^{2}+\frac{k}{R^{2}}\right]=\frac{b-r b^{\prime}}{2 r^{3}}-8 \pi P_{w}^{t}=m.
\end{eqnarray}

However, a new term arises corresponding to a linear potential. Note that $l$ and $m$ are constants. The separation constants $l$ and  $m$ are also determined by using the wormhole matter distribution. On the other hand, for the cosmological part, the conservation laws $T_{\nu ; \mu}^{\mu}=0$ give us the following equations:
\begin{equation} \label{ecc}
\dot{\rho}_{c}+3 H\left(\rho_{c}+P_{c}\right)=\frac{q}{k} \dot{R} R^{-3},
\end{equation}

where $q=l+3 m$. To investigate the cosmological part, we use the equation of state (EOS) of CG as follows \cite{Gorini:2004by}:

\begin{equation}
P_c=\frac{-A}{\rho_{c}} \label{DS}
,\end{equation} 
where $A$ is constant. The solution of the Eq. \ref{ecc} for $q=0$ by using the Eq. \ref{DS}, give us the cosmic density for Chaplygin gas $\rho_{\mathrm{c}}$ as follows
\begin{equation}
\rho_{c}=\sqrt{B+\frac{A}{R^{6}}}, \end{equation}
where $B$ is integration constant. For small $R$ ($\rho_c \sim R^{6} \ll A /B $), this solution reduces to $\rho_c \sim \frac{\sqrt{A}}{R^{3}}$. On the other hand, for large value of $R$, $\rho_c \sim \sqrt{B}, \tau_c \sim-\sqrt{B}$. Using the Eq.s \ref{S00}-\ref{S22} with $l=-m=0$, we find the following FRW equations:

\begin{eqnarray} \label{con1}
H^{2}=\left(\frac{\dot{R}}{R}\right)^{2}=\frac{8}{3} \pi \rho_{c}-\frac{k}{R^{2}}, \\
-2 \frac{\ddot{R}}{R}-\left(\frac{\dot{R}}{R}\right)^{2}=8 \pi P_{c}+\frac{k}{R^{2}}.
\end{eqnarray}

Then, using the cosmological matter distribution $\rho_c \sim \frac{\sqrt{A}}{R^{3}}$, and $k=0$, we find the scale factor $R(t)$ for small value of $R$
\begin{equation}
R(t) =\frac{\left(3 c+2 \sqrt{6 \pi } \sqrt[4]{A } t\right){}^{2/3}}{2^{2/3}},
\end{equation}  which is for a universe dominated by dust-like matter. We also calculate the scale factor R(t) for the large value of $R$, using the $\rho_c \sim \sqrt{B}$:

\begin{equation}
    R(t)=e^{2 \sqrt{\frac{2 \pi }{3}} \sqrt[4]{B} t},
\end{equation}
which is for an empty universe with a cosmological constant. 

In the next section, we will check the possibility of evolving topologically deformed wormhole supported various dark matter halos.

\section{Construction of Evolving Topologically Deformed Wormhole with Dark Matter Halo}
Now, we construct the evolving topologically deformed wormholes supported in the dark matter halo. For this purpose, we use the wormhole part of the Eq.s \ref{S00}-\ref{S22} by choosing $l=-3m$

\begin{eqnarray}\label{WS00}
\frac{b^{\prime}}{r^{2}}-8 \pi \rho_{w}+{\frac {-{\alpha}^{2}+1}{{r}^{2}{\alpha}^{2}}}=-3 m, \\
-\frac{b}{r^{3}}-8 \pi P_{w}^{r}-{\frac { \left( \alpha-1 \right)  \left( \alpha+1 \right)  \left( -k{
r}^{3}+r{R}^{2}-b+r \right) }{{r}^{2}{\alpha}^{2} \left( k{r}^{3}+b-r
 \right) }}=m, \label{WS11} \\
\frac{b-r b^{\prime}}{2 r^{3}}-8 \pi P_{w}^{t}=m.
\label{WS22}
\end{eqnarray}

From Eqs.~(\ref{WS00})-(\ref{WS22}) we obtain that
$\rho_w+P_{w}^{r}+2P_{w}^{t}=0$. We choose the one of the pressure in the form of $P_{w}^{r}=\omega_{r} \rho_{w}$, then the laterial pressure becomes barotropic $P_{w}^{t}=-\left(1+\omega_{r}\right) \rho_{w} / 2$. 
%%%%
 To check the maintenance of the wormhole with time, we study the embedding space in a flat three dimensional Euclidean space using the $t={\rm const}$ with equatorial plane $\theta=\pi/2$:
\begin{equation}
ds^2=d{\bar{z}}^2+d{\bar{r}}^2+{\bar{r}}^2\,\alpha^2 {d\varphi}^2\,.
\label{barredslice}
\end{equation}
Taking the relations: $\bar{r}={R(t)\,r}$, ${d\bar{r}}^2=R^2(t)\,{dr}^2$, and constant time slice \cite{Roman:1992xj} we  get

\begin{equation}\label{slice}
ds^2={R^2(t)\,{dr^2}\over{1-k r^{2}-\frac{b(r)}{r}}} + R^2(t)\,
r^2 \alpha^2 \,d\varphi^2\,.
\end{equation}
Then it is obtained that \cite{Morris:1988cz}:
\begin{equation}
{{d{\bar{z}}}\over{d{\bar{r}}}}
=\pm\left({{k\bar{r}^3+\bar{b}}\over{\bar{r}-k\bar{r}^3-\bar{b}(\bar{r})}}\right)^{1/2}
={{dz}\over{dr}} \,.
            \label{barredembedding}
\end{equation}
Eq. (\ref{barredembedding}) reduces to
\begin{eqnarray}
\bar{z}(\bar{r}) &=& \pm\int{{d\bar{r}}\left({{k\bar{r}^3+\bar{b}}\over{\bar{r}-k\bar{r}^3-\bar{b}(\bar{r})}}\right)^{1/2}}= \pm \, R(t)\,\int{\left({{kr^3+b}\over{r-kr^3-b(r)}}\right)^{1/2}} \,dr
	\nonumber \\
&=& \pm \,R(t)\,z(r)\,.
            \label{embed:relation}
\end{eqnarray}
The flare-out condition for the evolving wormhole become:
 $d\,^2{\bar{r}(\bar{z})}/d{\bar{z}}^2>0$ near the throat: \begin{equation}
{{d\,^2{\bar{r}(\bar{z})}}\over{d{\bar{z}}^2}}
       =\frac{1}{R(t)}\,{\left({{\bar{b}-{\bar{b}}'\bar{r}-2k\bar{r}^3}}\right)}
       =\frac{1}{R(t)}\,{{d\,^2r(z)}\over{dz^2}}>0\,.
       \label{barred:flareout}
\end{equation}
 where ${\bar{b}}'(\bar{r})={{d\bar{b}}/{d\bar{r}}}=b'(r)={{db}/{dr}}$. For the constant $R(t)=1$, the flare-out condition reduces to the static wormhole at throat: 
\begin{equation}
{{d\,^2{\bar{r}(\bar{z})}}\over{d{\bar{z}}^2}}
=\left({{\bar{b}-{\bar{b}}'\bar{r}-2k\bar{r}^3}}\right)>0\,.
         \label{barred:flareout2}
\end{equation}

\subsection {NFW dark matter profile}
 NFW density profile is used to study dark matter halo for galaxies and clusters. Here we find a wormhole solution in the NFW dark matter density profile, by using $\rho_w$: \cite{NFW}
 
\begin{equation}\label{E:rho}
\rho_{w}=\frac{\rho_{s}}{\frac{r}{R_{s}}\left(1+\frac{r}{R_{s}}\right)^{2}}, 
\end{equation}
where $R_s$ is known as radius of characteristic scale and $\rho_s$ is the dark matter density when the dark matter halo collapses. Using the Eq. \ref{E:rho} within the Eq. \ref{WS00}, and using the boundary condition for the wormhole $b(r_0)=r_0$ the shape function $b(r)$ is calculated as follows:

\begin{eqnarray} \label{w1}
   b(r)=\frac{8 \pi  R_{s}^4 \rho_{s} (r_{0}-r)}{(R_{s}+r_{0}) (R_{s}+r)}-8 \pi  R_{s}^3 \rho_{s} \ln (R_{s}+r_{0})+8 \pi  R_{s}^3 \rho_{s} \ln (R_{s}+r)+\frac{\alpha ^2 m \left(r_{0}^3-r^3\right)+r_{0}-r}{\alpha ^2}+r.
\end{eqnarray}

For the condition of $m=0$; shape function $b(r)$ reduces to 

\begin{equation} \label{w10}
b(r)=\frac{8 \pi  R_{s}^4 \rho_{s} (r_{0}-r)}{(R_{s}+r_{0}) (R_{s}+r)}-8 \pi  R_{s}^3 \rho_{s} \ln (R_{s}+r_{0})+8 \pi  R_{s}^3 \rho_{s} \ln (R_{s}+r)+\frac{r_{0}-r}{\alpha ^2}+r.
\end{equation}

\begin{figure}[ht!] 
   \centering 
    \includegraphics[scale=0.6]{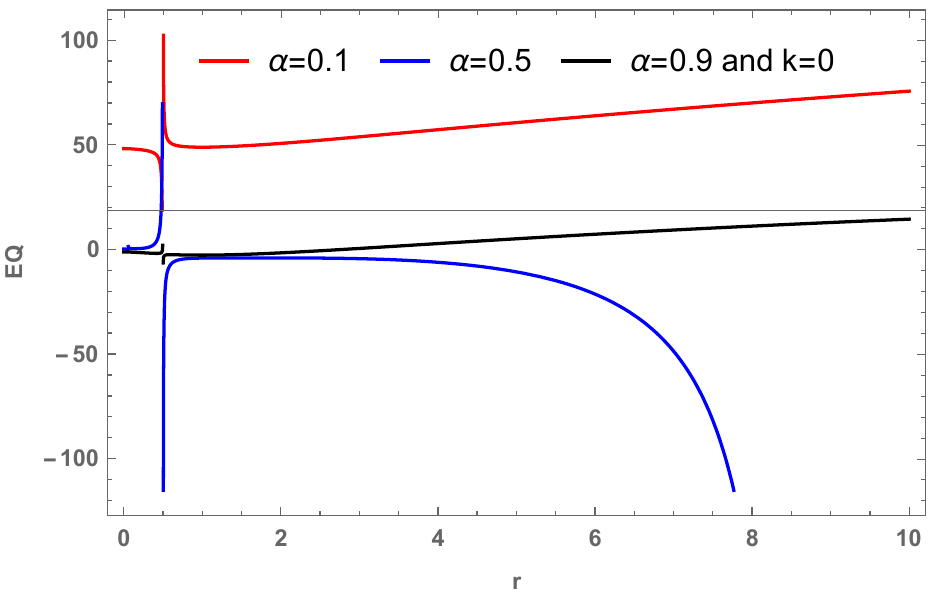}
    \caption{Plot of the flare-out equation $EQ$ vs $r$ which must be positive.}
    \label{wh1a}
\end{figure}

\begin{figure}[ht!]
   \centering
    \includegraphics[scale=0.6]{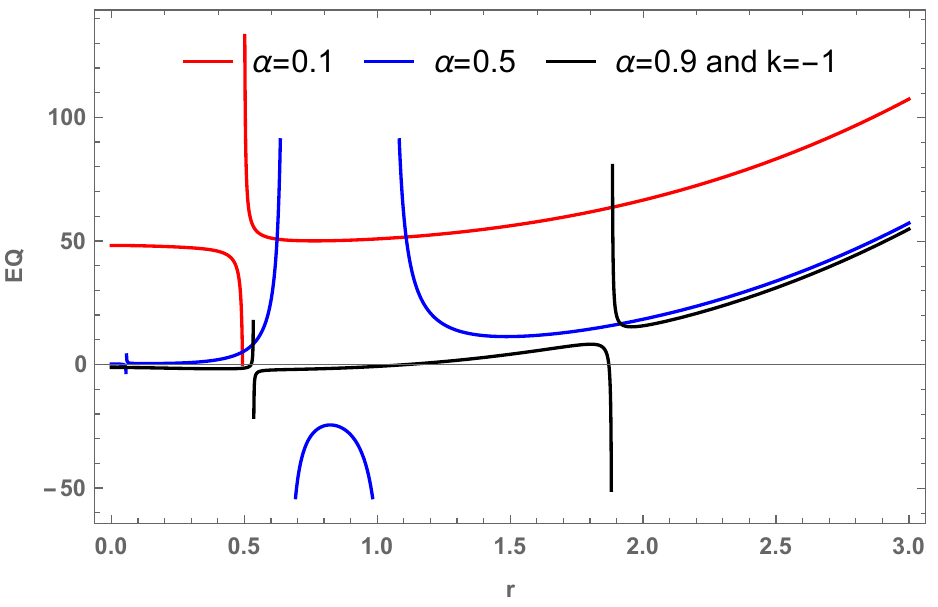}
    \caption{Plot of the flare-out equation $EQ$ vs $r$ which must be positive.}
    \label{wh1b}
\end{figure}

For constant $R=1$, we check the maintenance of the shape of the traversable wormhole, using the following flare-out equation which must be positive and derived by
violating the null energy condition (NEC) ($\rho+P^{r}\geqslant0$) via Eq.s \ref{4rGott}-\ref{4rGorr} 
\begin{equation} \label{eqq}
EQ=-\frac{r^2}{\alpha ^2 \left(b(r)+k r^3-r\right)}+\frac{r^2}{b(r)+k r^3-r}+b(r)-r b'(r)-2 k r^3>0,
\end{equation}

which is plotted in Fig.s \ref{wh1a} and \ref{wh1b}, showing that Eq. \ref{eqq} is not satisfied on the whole domain and bounded. Moreover,  the evolving  topologically deformed wormhole may be created with normal matter, however, Eq. \ref{eqq} will become negative and its lifetime can be said that limited. It is clearly shown in Figures that the topologically deformation parameter $\alpha$ change the stable region for the wormhole.

%moreover for $\alpha=1$ reduces to:
%\begin{equation}
 %    b(r)-r b'(r)-2k r^3>0.
%\end{equation}

%%%%%

%Then the NEC condition at the s the radius of the throat of the wormhole $r = r_0$, become

\subsection {Thomas-Fermi (TF) profile}
Here we construct a wormhole solution in the model of Bose-Einstein Condensation (BEC) dark matter (DM) which has more advantages on the small scales of galaxies. For this model, density profile  is \cite{Boehmer:2007um}

\begin{equation}
\rho_{\mathrm{TF}}=\rho_{s} \frac{\sin (k r)}{k r}
\end{equation}

where $\rho_s$ is the center density of Bose-Einstein Condensation dark matter halo, and $k = \pi/R$ is
the radius where the dark matter pressure and density vanish. Note that The BEC-DM model supports much less dark matter density in the center regions of galaxies than the NFW profile.

\begin{equation} \label{w2}
    b(r)= \frac{8 \rho_{s} R^2 \left(R \left(\sin \left(\frac{\pi  r}{R}\right)-\sin \left(\frac{\pi  r_{0}}{R}\right)\right)+\pi  r_{0} \cos \left(\frac{\pi  r_{0}}{R}\right)-\pi  r \cos \left(\frac{\pi  r}{R}\right)\right)}{\pi ^2}+r_{0}^3 m+\frac{r_{0}-r}{\alpha ^2}-m r^3+r
\end{equation}

\begin{figure}[ht!]
   \centering
    \includegraphics[scale=0.6]{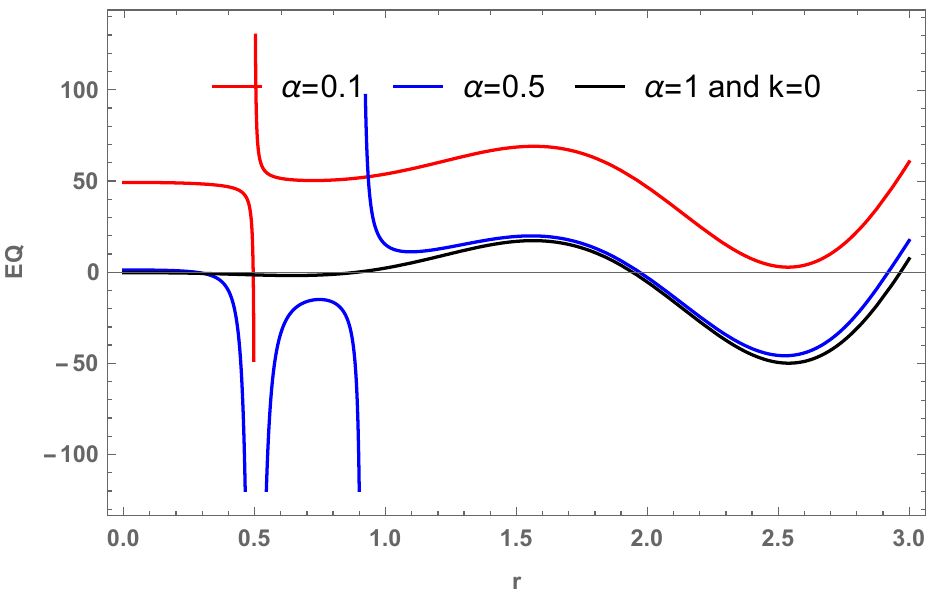}
    \caption{Plot of the flare-out equation $EQ$ vs $r$ which must be positive.}
    \label{w2a}
\end{figure}

\begin{figure}[ht!]
   \centering
    \includegraphics[scale=0.6]{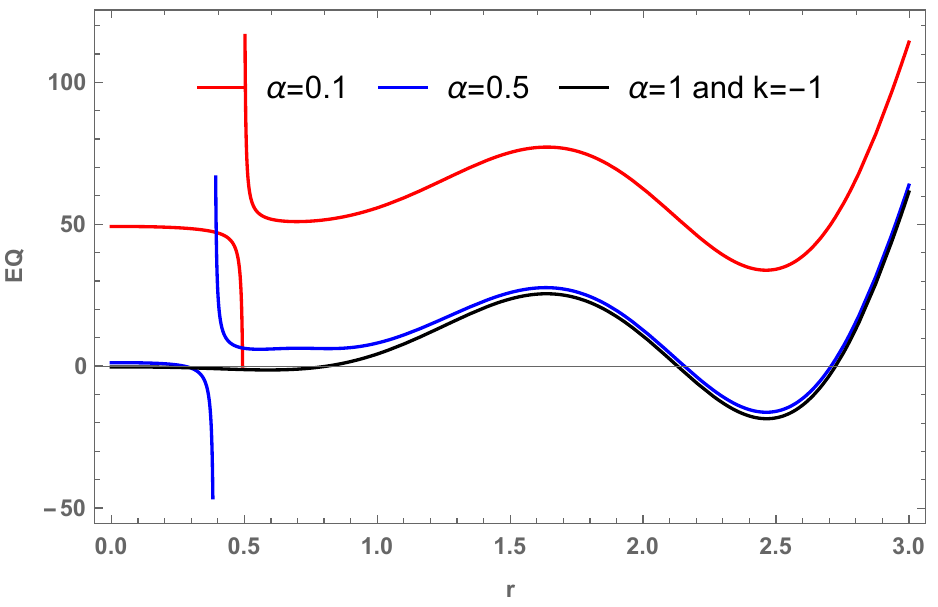}
    \caption{Plot of the flare-out equation $EQ$ vs $r$ which must be positive.}
    \label{w2b}
\end{figure}

For constant $R$, we check the maintenance of the shape of the traversable wormhole, using the following flare-out equation which must be positive
\begin{equation} \label{eqq2}
EQ=-\frac{r^2}{\alpha ^2 \left(b(r)+k r^3-r\right)}+\frac{r^2}{b(r)+k r^3-r}+b(r)-r b'(r)-2 k r^3>0.
\end{equation}
Note that it is plotted in Fig.s \ref{w2a} and \ref{w2b} showing that Eq. \ref{eqq2} is not satisfied on the whole domain and bounded. Moreover,  the evolving  topologically deformed wormhole may be created with normal matter, however, Eq. \ref{eqq2} will become negative and its lifetime can be said that limited.

\subsection{Wormhole supported Chaplygin gas and NFW dark matter profile}

Here, we combine the Chaplygin gas for the large value of $R$ with the NFW dark matter profile $\rho_w=\rho_c+\rho_{NFW}$ to obtain the shape function of the evolving topologically deformed wormholes.
The new combined energy density becomes:
\begin{equation}\label{E:rho22}
\rho_{w}=\frac{\rho_{s}}{\frac{r}{R_{s}}\left(1+\frac{r}{R_{s}}\right)^{2}}+\sqrt{B}, 
\end{equation}

then, the shape function is calculated as follows:
\begin{eqnarray} \label{w3}
b(r)&=&-\frac{8 \pi  R_{s}^4 \rho_{s}}{R_{s}+r_{0}}-\frac{8 \pi  R_{s}^4 \rho_{s} \log (R_{s}+r_{0})}{R_{s}+r_{0}}+\frac{8 \pi  R_{s}^4 \rho_{s}}{R_{s}+r}-\frac{8 \pi  R_{s}^3 \rho_{s} r_{0} \log (R_{s}+r_{0})}{R_{s}+r_{0}}+8 \pi  R_{s}^3 \rho_{s} \log (R_{s}+r)-\frac{8 \pi  \sqrt{B} r_{0}^4}{3 (R_{s}+r_{0})} \notag \\&-&\frac{8 \pi  R_{s} \sqrt{B} r_{0}^3}{3 (R_{s}+r_{0})}+\frac{r_{0}^4 m}{R_{s}+r_{0}}+\frac{R_{s} r_{0}^3 m}{R_{s}+r_{0}}+\frac{r_{0}^2}{\alpha ^2 (R_{s}+r_{0})}+\frac{R_{s} r_{0}}{\alpha ^2 (R_{s}+r_{0})}+\frac{8}{3} \pi  \sqrt{B} r^3 -m r^3-\frac{r}{\alpha ^2}+r. 
\end{eqnarray}

\begin{figure}[ht!]
   \centering
    \includegraphics[scale=0.6]{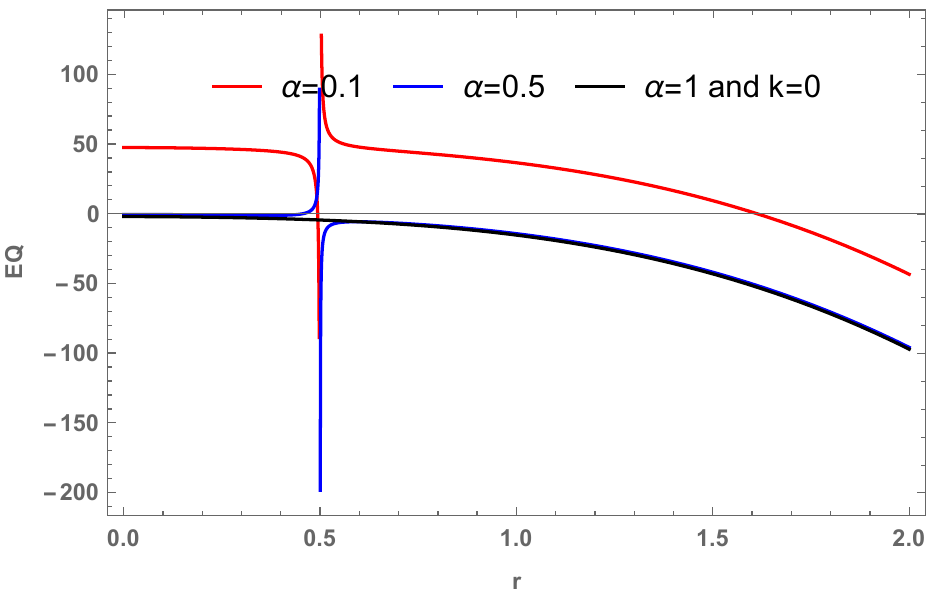}
    \caption{Plot of the flare-out equation $EQ$ vs $r$ which must be positive.}
    \label{wh3a}
\end{figure}

\begin{figure}[ht!]
   \centering
    \includegraphics[scale=0.6]{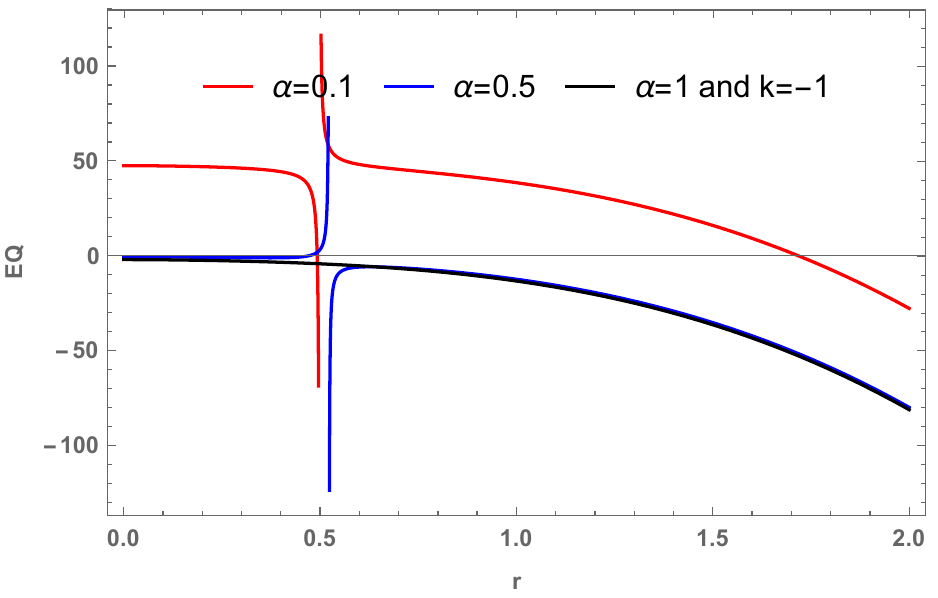}
    \caption{Plot of the flare-out equation $EQ$ vs $r$ which must be positive.}
    \label{wh3b}
\end{figure}

For constant $R$, we check the maintenance of the shape of the traversable wormhole, using the following flare-out equation which must be positive
\begin{equation} \label{eqqq}
EQ=-\frac{r^2}{\alpha ^2 \left(b(r)+k r^3-r\right)}+\frac{r^2}{b(r)+k r^3-r}+b(r)-r b'(r)-2 k r^3>0,
\end{equation}
which is plotted in Fig.s \ref{wh3a} and \ref{wh3b} showing that Eq. \ref{eqqq} is not satisfied on the whole domain and bounded. Moreover,  the evolving  topologically deformed wormhole may be created with normal matter, however, Eq. \ref{eqqq} will become negative and its lifetime can be said that limited.

\section{Conclusions}

In this paper, we have obtained the evolving topologically deformed wormhole supported in the dark matter halo and check its behavior of the wormhole in the inflation era. 
The solution of the field equations given in Eq.s \ref{4rGott}-\ref{4rGohhpp}, have been divided into a two independent system using the ansatz given in Eq.s~\ref{uno}-\ref{uno2}.  One of the Einstein field equations is for the static gravitational field on the other hand, second one is for the time dependent gravitational field. We have used the first part of the field equation to construct evolving topologically deformed wormhole by obtaining the shape function $b(r)$ for different models of dark matter halo, on the other hand, second equation which is the time dependent part (Friedmann-like with curvature $k$), have been used to study cosmological models within Chaplygin gas by obtaining the scale factor $R(t)$. To do so, we have used the Chaplygin gas as the equation of states in the cosmological part and three different dark matter models for the evolving topologically deformed wormhole part such as NFW dark matter profile, Thomas-Fermi profile and combination of the Chaplygin gas with NFW dark matter profile. Then the shape functions of the evolving wormhole are derived for these three models separately, and we have plotted their NEC conditions to show their maintenance. The effect of topologically deformation parameter $\alpha$ to present in the evolving wormhole metric is also explored. During the inflation era, the amount of the exotic matter needed is decreasing for the evolving wormhole as compared to the static wormhole. In the Fig.s \ref{wh1a} and \ref{wh1b}, we have plotted the $EQ$ versus ‘$r$’ with different values of the topologically deformation parameter $\alpha$ to show the effect of the $\alpha$ on the NEC. Second, we have constructed the evolving topologically deformed wormhole using the TF profile for BEC-DM model. The Fig.s \ref{w2a} and \ref{w2b}, show that NEC is not satisfied on the whole domain and bounded. Moreover, the evolving  topologically deformed wormhole may be created with normal matter, however, Eq. \ref{eqqq} will become negative and its lifetime can be said that limited. Last, we have combined the Chaplygin gas with NFW dark matter to obtain the shape function of the evolving topologically deformed wormhole. Here, we have showed that NEC is in general satisfied for the range of parameters shown in the Fig.s \ref{wh3a} and \ref{wh3b}. Our results have showed that the topologically deformation parameter $\alpha$ affects the stable region for the wormhole.

\end{document}